\documentclass{article}
\usepackage{setspace}
\usepackage{graphicx}
\usepackage{mathrsfs}
\usepackage{float}
\usepackage{array}

\usepackage{multicol}
\usepackage{amsmath}
\usepackage[colorlinks,linkcolor=black,filecolor=black,urlcolor=black,citecolor=black]{hyperref}
\usepackage{diagbox}
\numberwithin{equation}{section}
\usepackage{amsfonts}
\usepackage{amssymb,amsmath,cite}
\usepackage{amsthm}
\usepackage{mathrsfs}
\usepackage{indentfirst}
\usepackage{times}

\newcommand\be{\begin{equation}}
\newcommand\ee{\end{equation}}

\allowdisplaybreaks[2]

\newtheorem{theorem}{Theorem}
\newcommand\ber{\begin{eqnarray}}
\newcommand\eer{\end{eqnarray}}
\newcommand\berr{\begin{eqnarray*}}
\newcommand\eerr{\end{eqnarray*}}
\newcommand\bea{\begin{eqnarray}}
\newcommand\eea{\end{eqnarray}}

\newcommand{\bfR}{{\Bbb R}}

\newcommand{\x}{{\bf x}}

\newcommand{\vphi}{{\varphi
}}

\newcommand{\dd}{\mbox{d}}

\newcommand{\pa}{\partial}
\newcommand{\vep}{\varepsilon}

\newcommand{\nn}{\nonumber}

\newcommand\lb{\label}
\newcommand\eq{\eqref}
\newcommand\re{r_{\mbox{e}}}
\newcommand\rMBI{r_{\rm{MBI}}}\newcommand\rBI{r_{\rm{BI}}}

\begin{document}
\title{\Large The Maxwell--Born--Infeld Theory: Presence of Finite-Energy \\Electric Point Charge and Absence of Monopole and Dyon}
\author{{Tengyang Liu}$^{a,}$\footnote{Email address: tyl619@henu.edu.cn} { and Yisong Yang}$^{b,}$\footnote{Email address: yisongyang@nyu.edu}\\[2mm]{\it\small $^a$College of Mathematics and Statistics, Henan University}\\[2mm]{\it\small Kaifeng 475001, P.R.China}\\[2mm]{\it\small $^b$Courant Institute of Mathematical Sciences}\\{\it\small New York University}\\{\it\small New York, New York 10012, USA}}
\date{}

\maketitle
\begin{abstract}
\noindent We formulate a nonlinear electrodynamic theory which may be viewed as a weighted theory minimally interpolating the classical Maxwell and Born--Infeld theories. We show that, in
contrast to the Born--Infeld theory,  this new theory accommodates a finite-energy electric point charge, like that in the Born--Infeld theory, but does not accommodate a finite-energy magnetic point charge, known as the monopole, thereby exhibiting an electromagnetic asymmetry property, unlike that in the Born--Infeld theory. We estimate the radius of the electron within the formalism of such a theory. We also show that an electric point charge carries a finite energy in the Maxwell theory limit. Furthermore, we demonstrate that the theory does not accommodate a finite-energy monopole nor a
dyon either in its most general setting.\\




\end{abstract}

\section{Introduction}

\noindent Recall that the classical Maxwell theory of electromagnetism is formulated by an action density, $\mathcal{F}$, governing a real-valued gauge field $A_{\mu}$, with induced electromagnetic field tensor $F_{\mu\nu}$. In normalized units in empty space, it reads
\begin{align}\lb{1.1}
    \mathcal{F}=-\frac{1}{4}F_{\mu\nu}F^{\mu\nu},\ \ F_{\mu\nu}=\partial_{\mu}A_{\nu}-\partial_{\nu}A_{\mu}=
    \begin{pmatrix}
        0&-E^1&-E^2&-E^3\\
        E^1&0&-B^3&B^2\\
        E^2&B^3&0&-B^1\\
        E^3&-B^2&B^1&0
    \end{pmatrix},\ \ F^{\mu\nu}=\eta^{\mu\alpha}\eta^{\nu\gamma}F_{\alpha\gamma},
\end{align}
where $\eta^{\mu\nu}=\mbox{diag}\{1,-1,-1,-1\}$ is the flat Minkowski spacetime metric and $\textbf{E}=(E^1,E^2,E^3)$ and $\textbf{B}=(B^1,B^2,B^3)$ are the electric and magnetic fields, respectively, such that $\mathcal{F}=\frac{1}{2}\left(\textbf{E}^2-\textbf{B}^2\right)$. In order to resolve the problem in this theory that an electric point charge carries an infinite
energy, that is, the energy of  the electric field of a point charge given by Coulomb's law diverges,  Born \cite{B1,B2} and then Born and Infeld \cite{BI1,BI2} developed a nonlinear theory of electromagnetism which in its simplest form is given by the action density
\begin{align}
    \mathcal{L}(\mathcal{F})=\frac{1}{\beta}\left(1-\sqrt{1-2\beta \mathcal{F}}\right),\quad \beta>0,\label{BI-model}
\end{align}
where $\beta=\frac1{b^2}$ with $b>0$ being referred to as the Born coupling parameter. See (6.1) in Born \cite{B2} and (1.2) in Born and Infeld \cite{BI2}. The Born--Infeld theory is based on how classical mechanics and relativistic mechanics are related
through their respective action principles such that in the weak-field or small $\beta$ limit their theory \eq{BI-model} returns to the classical Maxwell theory \eq{1.1}, that is
\be\lb{1.3}
{\cal L}({\cal F})\approx {\cal F},\quad |\mathcal{F}|\ll 1\quad \mbox{or}\quad \beta\ll 1.
\ee
In contemporary research, the theory \eq{BI-model} has fruitfully been further extended to a general setting preserving the property \eq{1.3} with $\cal L({\cal F})$ 
simply being assumed to satisfy the imposed condition
\be\lb{1.4}
{\cal L}={\cal L}({\cal F}):\quad {\cal L}(0)=0,\quad {\cal L}'(0)=1.
\ee
See \cite{JHOR,Y-book,Y-survey} and references therein for the vast literature on this subject. In \cite{B1,B2,BI1,BI2}, besides being successful in achieving a finite-energy for an electric point charge,
an estimate for the electron radius, $r_0$, is also obtained to be
\be\lb{1.5}
r_0\approx2.28\times 10^{-13} \mbox{cm},
\ee
such that the Born parameter enjoys the estimated value
\be\lb{1.6}
b\approx 9.24\times 10^{15}\mbox{cm}^{-\frac12}\mbox{g}^{\frac12}\mbox{s}^{-1},
\ee
which is consistent with the presumption that $b$ is a very large quantity. Interestingly,  the Born--Infeld theory possesses the property that
both electric and magnetic point charges carry finite energies similar to what happens in the Maxwell theory that both electric and magnetic point charges carry an infinite energy. A magnetic 
point charge is called a monopole. Thus, in the sense of accommodating an electric point charge and a monopole, the Maxwell and Born--Infeld theories have the same property in view of
 energy consideration (the energy being either infinite or finite), thereby exhibiting an electromagnetic symmetry.

The concept of monopole was first advocated by P. Curie \cite{Curie} in 1894 and mathematically formulated by Dirac \cite{Dirac} within the Maxwell theory formalism in 1931.
However, despite of its prevalent popularity and valued implications in theoretical physics \cite{GO,MS,Pre,Wein,Yyellowbook,Zee}, monopole has never been observed
in laboratory, although its condensed-matter simulations have been developed and reported \cite{C,Gib,Gin,P,Q,R} recently. Thus, an important question arises whether one can come up with an appropriate 
electrodynamic theory of the type \eq{1.4} that accommodates a finite-energy electric point charge as in the Born--Infeld theory but excludes a finite-energy monopole, unlike that in
the Born--Infeld theory, hereby achieving an electromagnetic asymmetry. Such an asymmetry is established for all polynomial models \cite{Y1}. The interest of this result is that, generically
speaking \cite{Y-survey}, all generalized Born--Infeld theories enjoy the stated electromagnetic asymmetry, since the celebrated Stone--Weierstrass theorem \cite{Stone,Yosida} indicates that any nonlinear model may be
approximated by a sequence of polynomial models.

Therefore, it will be of fundamental importance to develop a minimally extended electrodynamic theory within which a finite-energy electric point charge is allowed but a monopole is prohibited.
In this work, we present such a theory to accomplish this goal.

The nonlinear electrodynamic theory to be presented takes the form of an interpolated theory between the Maxwell theory and the Born--Infeld theory.  In other words, this new theory
comes about as a weighted sum of the Maxwell and Born--Infeld theories. It is so structured that its Born--Infeld part serves to incorporate a finite-energy electric point charges, exactly as that in
the pure Born--Infeld theory, and that its Maxwell part plays the role of rejecting a finite-energy magnetic point charge, namely, a monopole.

In the next section, we formulate this new electrodynamic theory. In Section 3, we resolve the electrostatic constitutive equation associated with the theory. In particular, we solve the equations of motion and calculate the total energy of the solution describing an electric point charge. 
We show that the energy can be expressed explicitly in terms of the electric charge and effective radius of the point charge as in the classical Maxwell theory and the Born--Infeld theory.
As a by-product, we estimate the 
electron radius along the line of the calculation by Born and Infeld \cite{BI2}, in the entire parameter region.
In Section 4, we show that the energy of a monopole or a dyon
 diverges as a result of the Maxwell theory contribution, thereby establishing the property that the theory does not accommodate a monopole nor a dyon. In other words, concerning
a point charge, the finite-energy condition dictates that only an electric charge is allowed. In Section 5, we consider the electrostatic equations subject to an arbitrarily distributed electric
charge density and obtain the properties of the solutions. These solutions may represent a  system of multicentered electric charge distribution or a continuous distribution of the electric charge and are shown to give rise to a magnetic current in general situations in the context of Schwinger's covariant Maxwell equation formalism \cite{Sch1,Sch2,Sch3}. In Section 6, we draw conclusions.

\section{A new nonlinear electrodynamic theory}

\noindent Let $\kappa_1$ and $\kappa_2$ be two positive dimensionless weight parameters satisfying $\kappa_1+\kappa_2=1$.  
We propose a nonlinear electrodynamic theory governed by the following convex combination of the Maxwell and Born--Infeld theories:
\begin{align}
    \mathcal{L}(\mathcal{F})=\kappa_1\mathcal{F}+\frac{\kappa_2}{\beta}\left(1-\sqrt{1-2\beta \mathcal{F}}\right),\label{model}
\end{align}
which may naturally be referred to as a Maxwell--Born--Infeld theory. It is clear that \eq{model} fulfills the condition \eq{1.4} and minimally incorporates both the Maxwell theory \eq{1.1} and
the Born--Infeld theory \eq{BI-model}.
In the presence of an external current density $j_\mu$, the action density \eq{model} becomes
\begin{align}
    \mathcal{L}({\cal F})-A_{\mu}j^\mu=\kappa_1\mathcal{F}+\frac{\kappa_2}{\beta}\left(1-\sqrt{1-2\beta \mathcal{F}}\right)-A_\mu j^\mu,
\end{align}
whose  Euler--Lagrange equations with the variation of gauge field read
\begin{align}
    \partial_{\mu}P^{\mu\nu}=j^\nu,\ \ (P^{\mu\nu})=\left(\mathcal{L}^\prime(\mathcal{F})F^{\mu\nu}\right)=
    \begin{pmatrix}
        0&-D^1&-D^2&-D^3\\
        D^1&0&-H^3&H^2\\
        D^2&H^3&0&-H^1\\
        D^3&-H^2&H^1&0
    \end{pmatrix},\label{2.3}
\end{align}
where $\textbf{D}=(D^1,D^2,D^3)$ and $\textbf{H}=(H^1,H^2,H^3)$ are the electric displacement and magnetic intensity fields, respectively. See also \cite{BB}.
Thus, in light of \eq{1.1} and \eq{2.3}, the associated constitutive equations relating the fields ${\bf E},{\bf B}$ and ${\bf D},{\bf H}$ are
\bea
    &&\textbf{D}=\mathcal{L}^\prime(\mathcal{F})\textbf{E}=\left(\kappa_1+\frac{\kappa_2}{\sqrt{1-\beta({\bf E}^2-{\bf B}^2)}}\right){\bf E},\lb{2.4}\\
&&\textbf{H}=\mathcal{L}^\prime(\mathcal{F})\textbf{B}=\left(\kappa_1+\frac{\kappa_2}{\sqrt{1-\beta ({\bf E}^2-{\bf B}^2)}}\right){\bf B}.\label{2.5}
\eea
Besides, the Hamiltonian density of \eq{model} is given by \cite{Y-PRD,Y-AOP,Y-AOP2}
\begin{align}\lb{2.6}
    \mathcal{H}&=\mathcal{L}^\prime(\mathcal{F})\textbf{E}^2-\mathcal{L}(\mathcal{F})\notag\\
    &=\left(\kappa_1+\frac{\kappa_2}{\sqrt{1-\beta({\bf E}^2-{\bf B}^2)}}\right){\bf E}^2-\left(\frac{\kappa_1}2({\bf E}^2-{\bf B}^2)+\frac{\kappa_2}\beta\left(1-\sqrt{1-\beta({\bf E}^2-{\bf B}^2)}\right)\right)\nn\\
&=\frac{\kappa_1}2({\bf E}^2+{\bf B}^2)+\kappa_2\left(\frac{{\bf E}^2}{\sqrt{1-\beta({\bf E}^2-{\bf B}^2)}(1+\sqrt{1-\beta({\bf E}^2-{\bf B}^2)})}+\frac{{\bf B}^2}{1+\sqrt{1-\beta({\bf E}^2-{\bf B}^2)}} \right),
\end{align}
which is a weighted sum of the Hamiltonian densities of the Maxwell and Born--Infeld theories with the same weight distribution as those in the action densities as stated in \eq{model}, 
not surprisingly.
This expression will be useful when we calculate the energies of an electric point charge, a monopole, and a dyon.

\section{An electric point charge}\lb{sec3}
\noindent
We consider an electrostatic point charge. In this situation, we have ${\bf B}={\bf H}={\bf 0}$ and 
 $\mathcal{F}=\frac{1}{2}\textbf{E}^2$ so that the system of the equations \eq{2.4} and \eq{2.5} becomes a single one:
\begin{align}\lb{3.1}
    \textbf{D}=\left(\kappa_1+\frac{\kappa_2}{\sqrt{1-\beta\textbf{E}^2}}\right)\textbf{E}.
\end{align}
It is difficult to resolve ${\bf E}$ in terms of $\bf D$ explicitly. Here, instead,  we pursue the solution to \eq{3.1} implicitly. 

First, squaring both sides of \eq{3.1} to get
\be\lb{a3.2}
\beta {\bf D}^2=\left(\kappa_1+\frac{\kappa_2}{\sqrt{1-\beta\textbf{E}^2}}\right)^2\beta\textbf{E}^2\equiv g(\eta),\quad \eta=\beta{\bf E}^2.
\ee
Since $g'(\eta)>0$, we see that \eq{a3.2} can be inverted to give us 
\be\lb{a3.3}
\beta{\bf E}^2= h(\beta{\bf D}^2),
\ee
where $h$ is the inverse of $g$. Substituting \eq{a3.3} back to \eq{3.1}, we obtain the induced electric field $\bf E$ in an implicit but unique way in terms of the electric displacement field
$\bf D$:
\be\lb{a3.4}
{\bf E}=\frac{\bf D}{\left(\kappa_1+\frac{\kappa_2}{\sqrt{1-h(\beta{\bf D}^2)}}\right)}.
\ee
On the other hand, in the electrostatic situation, the Hamiltonian energy density \eq{2.6} becomes
\begin{align}\lb{3.5}
    \mathcal{H}({\bf E}^2)
    &=\frac{\kappa_1}{2}\textbf{E}^2+\frac{\kappa_2}{\beta}\left(\frac1{\sqrt{1-\beta\textbf{E}^2}}-1\right)\nn\\
&=\left(\frac{\kappa_1}2+\frac{\kappa_2}{\sqrt{1-\beta{\bf E}^2}(1+\sqrt{1-\beta{\bf E}^2})}\right){\bf E}^2,
\end{align}
such that the total associated energy reads
\bea\lb{3.6}
E&=&\frac1{\beta}\int\left(\frac{\kappa_1}2+\frac{\kappa_2}{\sqrt{1-\beta{\bf E}^2}(1+\sqrt{1-\beta{\bf E}^2})}\right)\beta{\bf E}^2\dd\x\nn\\
&=&\frac1{\beta}\int\left(\frac{\kappa_1}2+\frac{\kappa_2}{\sqrt{1-h(\beta{\bf D}^2)}(1+\sqrt{1-h(\beta{\bf D}^2})}\right)h(\beta{\bf D}^2)\dd\x,
\eea
in terms of \eq{a3.3}. 

Although this expression appears complicated, it enjoys an elegant form in the point charge situation. To appreciate this structure,
 consider an electric point charge $q>0$ placed at the origin of the space $\bfR^3$.  In such a situation, the electric displacement field satisfies the equation
\be
\nabla\cdot{\bf D}=4\pi\delta(\x),
\ee
 so that $\bf D$ obeys Coulomb's law:
\begin{align}
    \textbf{D}=\frac{q\textbf{x}}{r^3},\ \ r=|\textbf{x}|,\ \ {\bf x}\in\bfR^3,\ \ {\bf x}\neq{\bf 0}.\label{3.8}
\end{align}
We follow the formalism of Born--Infeld \cite{BI1,BI2} to use the parameter 
\be\lb{3.14}
a=\beta^{\frac14} q^{\frac12}
\ee
 to represent the effective radius of the finite-energy charged point particle. Inserting \eq{3.8} into \eq{3.6} and using \eq{3.14}, we have
\bea
\frac E{4\pi}&=&\frac{1}\beta \int_0^\infty\left(\frac{\kappa_1}2+\frac{\kappa_2}{\sqrt{1-h\left(\frac{a^4}{r^4}\right)}(1+\sqrt{1-h\left(\frac{a^4}{r^4}\right)})}\right)h\left(\frac{a^4}{r^4}\right)r^2\dd r\nn\\
&=&\frac{q^2}a \int_0^\infty\left(\frac{\kappa_1}2+\frac{\kappa_2}{\sqrt{1-h\left(\frac{1}{x^4}\right)}(1+\sqrt{1-h\left(\frac{1}{x^4}\right)})}\right)h\left(\frac{1}{x^4}\right)x^2\dd x.\label{3.34}
\eea
In other words, the energy $E$ separates itself into the form
\be\lb{3.28}
\frac{E}{4\pi}=\frac{q^2}a H (\kappa_1,\kappa_2),
\ee
where $H$ is a model-dependent function depending only on $\kappa_1$ and $\kappa_2$ but not on $q$ and $a$. Thus it is dimensionless and may be referred to as the normalized energy
in our context here. In the limit $\kappa_1=0$ and $\kappa_2=1$, we have
\be\lb{3.12}
h(t)=\frac t{1+t}.
\ee
Substituting \eq{3.12} into \eq{3.34} with $\kappa_1=0$ and $\kappa_2=1$, we have the exact expression
\be\lb{a3.13}
H(0,1)=\left(\frac{\pi^{\frac32}}{3\Gamma\left(\frac34\right)^2}\right)\frac{ q^2}a\approx1.2361\frac{q^2}{a},
\ee
which is the classical result obtained in Born--Infeld \cite{BI1,BI2}. The expression \eq{3.28} is also in line of the classical formula
\cite{Gri,Jackson,Y-book}:
\be\lb{aa3.20}
E=\left(\frac35\right)\frac{q^2}{R},
\ee 
 giving the total self energy of a uniformly charged ball of radius $R$ and charge $q$, following Coulomb's law.

We now show how to calculate \eq{3.34}  explicitly and exactly despite its difficult appearance.

In fact, with \eq{3.8}, we can rewrite \eq{a3.2} as
\be
\frac1{x^4}=\left(\kappa_1+\frac{\kappa_2}{\sqrt{1-\eta}}\right)^2 \eta,\quad\eta=\beta{\bf E}^2=h\left(\frac1{x^4}\right),\quad x=\frac ra.
\ee
Thus
\be\lb{7.2}
x=\left(\kappa_1+\frac{\kappa_2}{\sqrt{1-\eta}}\right)^{-\frac12}\eta^{-\frac14},
\ee
which gives us the asymptotic relations for the pair $x$ and $\eta$ in the following correspondent ways:
\be\lb{7.3}
x=0,\quad \eta=1;\quad x\to\infty,\quad\eta=0.
\ee
Note that this correspondence of the asymptotics, \eq{7.3}, is valid for $\kappa_2>0$ but invalid for $\kappa_2=0$. That is, our formalism is nonlinear which does not
return to its linear theory limit or the classical Maxwell theory with setting $\kappa_2=0$.
In view of \eq{7.2} and \eq{7.3}, we see that \eq{3.34} leads to
\bea\lb{a3.18}
&&H (\kappa_1,\kappa_2)=\int_0^\infty\left(\frac{\kappa_1}2+\frac{\kappa_2}{\sqrt{1-\eta}(1+\sqrt{1-\eta})}\right)\eta x^2\dd x\nn\\
&&=\frac14\int_0^1\left(\frac{\kappa_1}2+\frac{\kappa_2}{\sqrt{1-\eta}(1+\sqrt{1-\eta})}\right)\left(\kappa_1+\frac{\kappa_2}{\sqrt{1-\eta}}\right)^{-\frac52}\eta^{-\frac34}
\left(\kappa_1+\frac{\kappa_2}{(1-\eta)^{\frac32}}\right)\,\dd\eta,\quad\quad
\eea
explicitly and exactly as anticipated.

We then delve on some real-data calculation. 

Recall that, using the known data on the electron rest mass $m$ to replace the total energy $E$ and the electron charge 
$e$ to replace the charge $q$ in \eq{aa3.20}, we may obtain the classical electron radius \cite{Gri,Y-book,Young}:
\be\lb{a3.19}
\re=\left(\frac35\right)\frac{e^2}{m}\approx 1.1067\times 10^{-13} \quad\mbox{cm}.
\ee
Thus, replacing the left-hand side of  \eq{3.28} by the electron mass $m$ and $q$ by the electron charge $e$ following Born and Infeld \cite{BI2}, we
arrive at the following formula for the electron radius for the Maxwell--Born--Infeld theory \eq{model}:
\be\lb{a3.20}
\rMBI=a=\frac{e^2}m H(\kappa_1,\kappa_2)\equiv \rMBI(\kappa_1,\kappa_2).
\ee
Therefore, inserting \eq{a3.13} into \eq{a3.20} and using \eq{a3.19}, we recover the result of Born and Infeld \cite{BI1,BI2} for the electron radius:
\bea\lb{3.17}
\rBI&=&\rMBI(0,1)=\left(\frac{5\pi^{\frac32}}{9\Gamma\left(\frac34\right)^2}\right)\re\nn\\
&\equiv& r_0 \approx 2.06\,\re,
\eea
giving rise to \eq{1.5} and the subsequent estimate \eq{1.6}.

As illustrations and for interest, we present a few cases based on the numerical integration of \eq{a3.18}. First, here are the ``midway" results for the 
normalized energy and electron radius:
\be
 H\left(\frac12,\frac12\right)=1.27275,\ \ \rMBI\left(\frac12,\frac12\right)=2.12125\,\re,
\ee
both are not far from those of Born--Infeld \cite{BI1,BI2}. In Table \ref{T1}, we list a few other cases.
\begin{table}[htbp]
\centering
\normalsize
\renewcommand{\arraystretch}{2.2}
\setlength{\tabcolsep}{4pt}
\centering
\begin{tabular}{|c|c|c|c|c|c|c|}
\hline
{$\kappa_1,\kappa_2$}&$\frac14,\frac{3}{4}$&$\frac15,\frac{4}{5}$&$\frac1{10},\frac{9}{10}$&$\frac{1}{15},\frac{14}{15}$&$\frac1{30},\frac{29}{30}$&$0,1$\\
\hline
$H$&$1.2528875$&$1.249315$&$1.242491$&$1.240304$&$1.238158$&$1.2361$\\
\hline
$\frac{r_{\rm{MBI}}}{\re}$&$2.08815$&$2.08219$&$2.07082$&$2.06718$&$2.06360$&$2.06017$\\
\hline
\end{tabular}\caption{Numerical results concerning normalized energy and the ratio of the Maxwell--Born--Infeld theory radius and the classical Maxwell theory radius of the electron.}\lb{T1}
\end{table}

These results indicate that both the normalized energy and electron radius slowly decrease to their Born--Infeld values as $\kappa_2$ reaches its limiting value $\kappa_2=1$. Below, we
prove this property.

For convenience, we set $\kappa_2=\tau$ and rewrite the dependence of quantities $\eta$ in \eq{a3.2} and $\cal H$ in \eq{3.5} as $\eta=\eta(\tau)$ and ${\cal H}={\cal H}(\tau)$,
respectively. Thus, we have
\bea
\beta {\bf D}^2&=&\left(1-\tau+\frac{\tau}{\sqrt{1-\eta(\tau)}}\right)^2\eta(\tau).\lb{a3.23}\\
{\cal H}(\tau)&=&\frac1\beta\left(\frac{1-\tau}2+\frac{\tau}{\sqrt{1-\eta(\tau)}(1+\sqrt{1-\eta(\tau)})}\right)\eta(\tau).\lb{a3.24}
\eea
Differentiating \eq{a3.23} with respect to $\tau$ and noting that the left-hand side of \eq{a3.23} does not depend on $\tau$, we obtain
\be\lb{a3.25}
\eta'(\tau)=\frac{2\eta(\sqrt{1-\eta}-1)}{(1-\tau)\sqrt{1-\eta}+\tau+\frac{\tau\eta}{1-\eta}}.
\ee
Then, differentiating \eq{a3.24} with respect to $\tau$ again and using \eq{a3.25}, we find after some lengthy algebra the simple result 
\be
{\cal H}'(\tau)=-\frac{\eta^2}{2\beta (1+\sqrt{1-\eta})^2}.\lb{a3.26}
\ee
That is, we have shown that the Hamiltonian energy density is a monotone decreasing function of $\tau=\kappa_2$ and that such a property is universally
valid regardless how the electric field and charge source are distributed.

Moreover, combining \eq{a3.25} and \eq{a3.26}, we have
\be
{\cal H}''(\tau)=\frac{\eta^2\sqrt{1-\eta}(1-\sqrt{1-\eta})}{\beta(1+\sqrt{1-\eta})((1-\eta)(1-\tau)\sqrt{1-\eta}+\tau)},
\ee
which indicates that the Hamiltonian energy density $\cal H$ is also strictly convex with respect to $\tau=\kappa_2$.

\begin{figure}[htbp]
    \centering
    \includegraphics[width=0.5\linewidth]{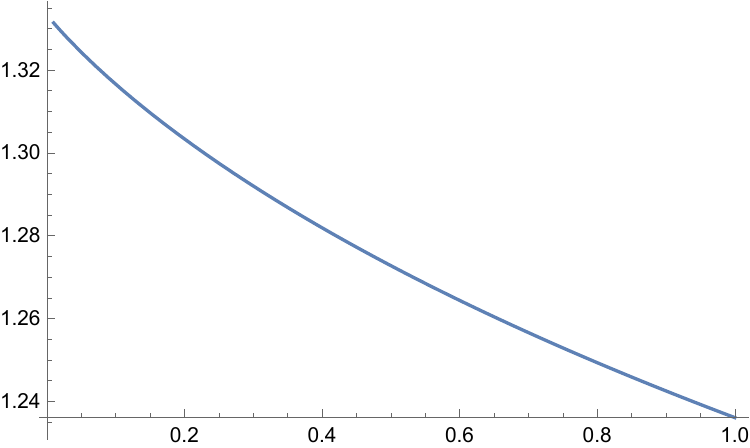}
    \caption{\normalsize{The normalized energy $H(\kappa_1,\kappa_2)$ plotted over the full parameter range $0<\kappa_2\leq 1$.}}
    \label{Fig1}
\end{figure}

In Figure \ref{Fig1}, we present the full curve of the normalized energy \eq{a3.18} based on numerical integration which confirms our mathematical results that
the energy density decreases and is concave over the full parameter range $0<\kappa_2\leq1$.

Of particular interest is what happens when $\kappa_2\to0$ (hence $\kappa_1\to 1$), which may be regarded as taking a Maxwell theory limit.  In Table \ref{T2}, we present some numerical
integration results showing the limit
\be\lb{aa3.28}
\lim_{\kappa_2\to0} H(\kappa_1,\kappa_2)=\frac43,
\ee
for the normalized energy. In consequence of this result and the expression \eq{3.28}, we arrive at the following formula for the ratio of the electron radius in the Maxwell theory limit against that in the classical theory:
\be
\lim_{\kappa_2\to0}\frac{\rMBI(\kappa_1,\kappa_2)}{\re}=\frac{20}9.
\ee

\begin{table}[htbp]
\centering
\normalsize
\renewcommand{\arraystretch}{2.2}
\setlength{\tabcolsep}{4pt}
\centering
\begin{tabular}{|c|c|c|c|c|c|c|c|}
\hline
{$\kappa_1,\kappa_2$}&$\frac34,\frac{1}{4}$&$\frac45,\frac{1}{5}$&$\frac9{10},\frac{1}{10}$&$\frac{19}{20},\frac{1}{20}$&$\frac{99}{10^2},\frac{1}{10^2}$&$\frac{9999}{10^4},\frac{1}{10^4}$&$\frac{999999}{10^6},\frac{1}{10^6}$\\
\hline
$H$&$1.2974825$&$1.3033526$&$1.316581$&$1.3242841$&$1.3313474$&$1.3333126$&$1.3333333$\\
\hline
$\frac{\rMBI}{\re}$&$2.16247$&$2.17225$&$2.1943$&$2.20714$&$2.21891$&$2.22219$&$2.22222$\\
\hline
\end{tabular}\caption{A study of the Maxwell theory limit with $\kappa_1\to1$ (hence $\kappa_2\to0$).}\lb{T2}
\end{table}

Interestingly, if we rewrite the normalized energy \eq{a3.18} as
\be
H(\kappa_1,\kappa_2)=\int_0^1 h(\eta;\kappa_1,\kappa_2)\,\dd\eta,
\ee
then we see that there holds the strict Fatou's inequality \cite{W}
\be\lb{aa3.31}
\int_0^1\lim_{\kappa_2\to0}h(\eta;\kappa_1,\kappa_2)\,\dd\eta=\frac18\int_0^1\eta^{-\frac34}\,\dd\eta=\frac12<\lim_{\kappa_2\to0}\int_0^1 h(\eta;\kappa_1,\kappa_2)\,\dd\eta =\frac43.
\ee
This property indicates that the analytic behavior of $h(\eta;\kappa_1,\kappa_2)$ as $\kappa_2\to 0$, although monotone, is rather subtle. Nevertheless, the property \eq{aa3.28} enables us to talk about
a finite-energy electric point charge, in particular, a classical point-charge electron, in the Maxwell theory limit of the Maxwell--Born--Infeld theory \eq{model}, which is interesting and in fact quite surprising. (The strict inequality \eq{aa3.31} subtly indicates that there is a lack of a dominant function for us to establish the existence of a finite limit in \eq{aa3.31} or \eq{aa3.28}
directly. Indirectly, however, \eq{a3.26} implies that $H(\tau)\equiv H(1-\tau,\tau)$ is a decreasing function of $\tau=\kappa_2$ such that the limit on the left-hand side of \eq{aa3.28}
exists and is positive, although it could be infinite. Numerical integration of \eq{a3.18} then shows that this limit is actually finite and valued at $\frac43$ as stated in \eq{aa3.28}. )

\section{Absence of finite-energy monopole and dyon across the theory}

\noindent Consider a point monopole of magnetic charge $g>0$ placed at the origin of $\bfR^3$, or more generally a point dyon placed at the same spot of electric charge $q\geq0$ and magnetic
charge $g>0$. Then its electric displacement field $\bf D$ and magnetic field $\bf B$ are governed by the point source equations
\be\lb{4.1}
\nabla \cdot {\bf D}=4\pi q\delta({\bf x}),\quad \nabla\cdot{\bf B}=4\pi g\delta({\bf x}),
\ee
where $\delta({\bf x})$ is the Dirac distribution concentrated at the origin ${\bf x}={\bf 0}$. Thus $\bf D$ is as given in \eq{3.8} and $\bf B$ is given by the same expression. That is,
\begin{align}
    \textbf{B}=\frac{g\textbf{x}}{r^3},\ \ r=|\textbf{x}|,\ \ {\bf x}\in\bfR^3,\ \ {\bf x}\neq{\bf 0}.\label{4.2}
\end{align}
On the other hand, in view of \eq{2.6}, we have
\be\lb{4.3}
{\cal H}\geq\frac{\kappa_1}2{\bf B}^2,
\ee
which leads to the divergence of the associated energy whenever $\kappa_1\neq0$ since $\bf B$ is given by \eq{4.2}, thereby establishing the exclusion of a monopole or dyon
with observation of the finite-energy condition. 

We next show that the same monopole-dyon exclusion property holds in the  Maxwell--Born--Infeld theory governed by the action density \eq{model} where  $\cal F$
is expanded to include
an electromagnetic self-interaction term:
\be\lb{4.4}
{\cal F}=-\frac{1}{4}F_{\mu\nu}F^{\mu\nu}+\frac{\kappa^2}{32}\left(F_{\mu\nu}\tilde{F}^{\mu\nu}\right)^2,
\ee
where $\kappa>0$ is taken to be an independent coupling parameter such that the limit $\kappa=0$ returns to the original model \eq{model} with \eq{1.1},
$\tilde{F}^{\mu\nu}=\frac12\vep^{\mu\nu\mu'\nu'}F_{\mu'\nu'}$ is the Hodge dual of $F_{\mu\nu}$, and the case $\kappa=b^{-1}$
is studied in \cite{BI2} based on a geometric invariance consideration. See (2.11) in Born and Infeld \cite{BI2}.  In the static situation, \eq{4.4} becomes 
\be\lb{4.5}
{\cal F}=\frac12({\bf E}^2-{\bf B}^2)+\frac{\kappa^2}2({\bf E}\cdot{\bf B})^2,
\ee
which clearly indicates that the presence of the parameter $\kappa$ switches on an interaction strength of the underlying electric and magnetic fields.
For a generalized electrodynamic theory governed by an action density ${\cal L}({\cal F})$ where $\cal F$ is defined in \eq{4.4}, the associated energy-momentum tensor $T_{\mu\nu}$
reads \cite{Y-PRD}:
\be\lb{4.6}
T_{\mu\nu}=
-{\cal L}'({\cal F})\left(F_{\mu\mu'}g^{\mu'\nu'}F_{\nu\nu'}-\frac{\kappa^2}4(F_{\mu'\nu'}\tilde{F}^{\mu'\nu'})F_{\mu\mu''}g^{\mu''\nu''}\tilde{F}_{\nu\nu''}\right)-g_{\mu\nu}{\cal L}({\cal F}).
\ee
Thus the Hamiltonian energy density ${\cal H}=T_{00}$ is found to be
\be\lb{4.7}
{\cal H}={\cal L}'({\cal F})({\bf E}^2+\kappa^2({\bf E}\cdot{\bf B})^2)-{\cal L}({\cal F}).
\ee
Substituting \eq{model} with \eq{4.5} into \eq{4.7}, we have
\be\lb{4.8}
{\cal H}=\kappa_1\left(\frac12({\bf E}^2+{\bf B}^2)+\frac{\kappa^2}2({\bf E}\cdot{\bf B})^2\right)+\kappa_2\left(\frac{{\bf E}^2+\kappa^2({\bf E}\cdot{\bf B})^2}{\sqrt{1-2\beta
{\cal F}}(1+\sqrt{1-2\beta{\cal F}})}+\frac{{\bf B}^2}{1+\sqrt{1-2\beta {\cal F}}}\right),
\ee
which extends the expression \eq{2.6}. In particular, \eq{4.3} still holds, which proves the same energy divergence of a monopole or dyon in view of \eq{4.2} provided that $\kappa_1>0$.

It will be of interest to write down the full governing equation of motion for the theory \eq{model} where $\cal F$ is as given in \eq{4.5}:
\bea
    &&\textbf{D}=\left(\kappa_1+\frac{\kappa_2}{\sqrt{1-2\beta{\cal F}}}\right)({\bf E}+\kappa^2({\bf E}\cdot{\bf B}){\bf B}),\lb{4.9}\\
&&\textbf{H}=\left(\kappa_1+\frac{\kappa_2}{\sqrt{1-2\beta {\cal F}}}\right)({\bf B}-\kappa^2({\bf E}\cdot{\bf B}){\bf E}).\label{4.10}
\eea
To maintain a finite energy for a point charge, we have to switch off the magnetic field by setting $g=0$ in \eq{4.1} which results in ${\bf B}=0$. Thus the system of the equations \eq{4.9}
and \eq{4.10} is reduced into the single one \eq{3.1} which has been solved by the formula \eq{a3.4}. In other words, we have obtained the most general finite-energy dyonic point charge
solution to \eq{4.9} and \eq{4.10}, which can only be purely electric with $q>0$ and $g=0$ in \eq{4.1}, which is the solution \eq{a3.4} to the electrostatic equation \eq{3.1}.

Note that, if we use ${\cal F}_{\rm{M}}$ and ${\cal F}_{\rm{BI}}$ to denote the quantity $\cal F$ given in \eq{1.1} and \eq{4.4} to label their Maxwell and Born--Infeld
theory origins, respectively,  then the interpolated model of our interest will be governed by the action density
\be\lb{4.11}
{\cal L}=\kappa_1{\cal F}_{\rm{M}}+\frac{\kappa_2}{\beta}\left(1-\sqrt{1-2\beta \mathcal{F}_{\rm{BI}}}\right).
\ee
In this situation the associated Hamiltonian energy density assumes the form
\be
{\cal H}=\frac{\kappa_1}2({\bf E}^2+{\bf B}^2)+\kappa_2\left(\frac{{\bf E}^2+\kappa^2({\bf E}\cdot{\bf B})^2}{\sqrt{1-2\beta
{\cal F}_{\rm{BI}}}(1+\sqrt{1-2\beta{\cal F}_{\rm{BI}}})}+\frac{{\bf B}^2}{1+\sqrt{1-2\beta {\cal F}_{\rm{BI}}}}\right),
\ee
which may also be derived by setting $\kappa$ to zero in the first term in \eq{4.8} with $\kappa_1$ as its coefficient. Therefore we conclude again that the interpolated theory \eq{4.11}
will not allow a finite-energy magnetic point charge. That is, no monopole nor dyon can be accommodated in this theory either but an electric point charge, whose construction returns to  that carried out in Section \ref{sec3}. Besides, for \eq{4.11}, the full constitutive equations are
\bea
    &&\textbf{D}=\kappa_1{\bf E}+\frac{\kappa_2}{\sqrt{1-2\beta{\cal F}_{\rm{BI}}}}({\bf E}+\kappa^2({\bf E}\cdot{\bf B}){\bf B}),\lb{4.13}\\
&&\textbf{H}=\kappa_1{\bf B}+\frac{\kappa_2}{\sqrt{1-2\beta {\cal F}_{\rm{BI}}}}({\bf B}-\kappa^2({\bf E}\cdot{\bf B}){\bf E}),\label{4.14}
\eea
following the additivity structure of the underlying Lagrangian field-theoretical formalism.

\section{Electrostatic solution in the general setting}

\noindent We emphasize that the electrostatic solution construction \eq{a3.4}, valid for the full system of the Maxwell--Born--Infeld equations,
\eq{4.9} and \eq{4.10}, and \eq{4.13} and \eq{4.14}, where ${\bf B}={\bf 0}$, does not require that the electric displacement field $\bf D$ be that given by a single electric point charge, defined by \eq{3.8}. Rather, it can be
induced from an arbitrarily distributed external electric charge density, say $\rho_e(\x)$, such that 
$\bf D$ satisfies the Poisson equation
\be
\nabla\cdot{\bf D}=\rho_e(\x).
\ee
For example, $\rho_e(\x)$ may be taken to represent a multiply distributed system of point charges of the form
\be\lb{5.2}
\rho_e(\x)=4\pi\sum_{i=1}^N q_i\delta(\x-\x_i),
\ee
or that of a continuously distributed external charge situation. Furthermore, the solution so-constructed enjoys the property that the total induced free electric charge
$q_{\rm{free}}$ coincides with
the prescribed one $q$,  
\be\lb{5.4}
q=\frac1{4\pi}\int_{\bfR^3}\nabla\cdot{\bf D}\,\dd\x=\frac1{4\pi}\int_{\bfR^3}\rho_e(\x)\,\dd\x=\frac1{4\pi}\lim_{R\to\infty}\int_{|{\bf x}|=R }{\bf D}\cdot\dd{\bf S},
\ee
provided that $\bf D$ is Coulomb-like at infinity. That is,
\be\lb{5.3}
{\bf D}=\mbox{O}(|\x|^{-2}),\quad |\x|\gg1,
\ee
which is consistent with the assumption that the total prescribed electric charge $q$,
 is finite due to the divergence theorem. In particular, in the situation \eq{5.2}, we have
\be
q=\sum_{i=1}^N q_i.
\ee

In fact, the relation \eq{3.1} implies ${\bf E}\to{\bf0}$ as ${\bf D}\to{\bf0}$. Applying this property in \eq{3.1} and using \eq{5.3}, we get
\be\lb{3.23}
{\bf E}-{\bf D}=\mbox{O}(r^{-6}),\quad r=|{\bf x}|\gg 1.
\ee
Thus the total free electric charge carried by the electric field $\bf E$ is
\bea
    q_{\rm{free}}=\frac1{4\pi}\int_{\mathbb{R}^3}\nabla\cdot {\bf E}\, \dd {\bf x}
=\frac1{4\pi}\lim_{R\to\infty}\int_{|{\bf x}|=R }{\bf E}\cdot\dd{\bf S}
=\frac1{4\pi}\lim_{R\to\infty}\int_{|{\bf x}|=R }{\bf D}\cdot\dd{\bf S}=q, \label{3.24}
\eea
in view of \eq{5.4}. This result coincides that obtained by Born and Infeld in \cite{BI1,BI2} for their classical theory
corresponding to the limiting situation $\kappa_1=0, \kappa_2=1$ in our theory \eq{model}, for a point charge.  

Furthermore, in view of the formula $\nabla\times (\vphi{\bf F})=\nabla\vphi\times{\bf F}+\vphi \nabla\times{\bf F}$ for a scalar field $\vphi$ and a vector field $\bf F$, we see that \eq{a3.4} gives us
\be\lb{5.5}
\nabla\times{\bf E}=\frac{\nabla\times{\bf D}}{\left(\kappa_1+\frac{\kappa_2}{\sqrt{1-h(\beta{\bf D}^2)}}\right)}
-\frac{\beta \kappa_2 h'(\beta{\bf D}^2)({\bf D}\times \nabla{\bf D}^2)}{2\left(\kappa_1+\frac{\kappa_2}{\sqrt{1-h(\beta{\bf D}^2)}}\right)^2(1-h(\beta{\bf D}^2))^{\frac32}}.
\ee
This calculation enables us to examine the physical meaning of our solution in the context of Schwinger's covariant Maxwell equations \cite{Sch1,Sch2,Sch3}:
\be\lb{5.6}
\frac{\pa {\bf B}}{\pa t}+\nabla\times{\bf E}=-{\bf j}_m,\quad \frac{\pa {\bf D}}{\pa t}-\nabla\times{\bf H}=-{\bf j}_e,\quad \nabla\cdot{\bf D}=\rho_e,\quad \nabla\cdot{\bf B}=\rho_m,
\ee
where the pairs ${\bf j}_e$ and ${\bf j}_m$, and $\rho_e$ and $\rho_m$, are the electric and magnetic current and charge densities, respectively, arising as a result of 
electromagnetic duality. Thus, \eq{a3.4} solves the electrostatic situation of the system \eq{5.6} corresponding to the absence of its magnetic field sector. That is, the system now reduces
itself into
\be\lb{5.7}
\nabla\times{\bf E}=-{\bf j}_m,\quad \nabla\cdot{\bf D}=\rho_e.
\ee
Hence, we see that, if $\rho_e$ is radially symmetric, so is $\bf D$, which leads to the radial symmetry of $\bf E$ in view of \eq{a3.4}. As a consequence
of this, we see that the right-hand side of \eq{5.5}
vanishes, which leads to ${\bf j}_m={\bf0}$ in \eq{5.7}. In other words, a singly-centered electric charge distribution is purely electric so that the system does not maintain a magnetic
current. On the other hand, however, if the electric charge distribution is not singly centered, then the induced electric displacement field will not be radially symmetric and that the associated
electric field may not be solenoidal. That is, the system requires the presence of a magnetic current to maintain the static balance of a non-singly-centered electric charge distribution,
as in \cite{Y-AOP}.

Next, we show that, although an explicit expression for \eq{a3.4} is not available, we can resort to comparison estimates to obtain some explicit information. For this, 
we note that  we can square \eq{3.1} and use 
\be
\frac{\kappa_2}{\sqrt{1-\beta\textbf{E}^2}}\leq \kappa_1+\frac{\kappa_2}{\sqrt{1-\beta\textbf{E}^2}}\leq\frac{1}{\sqrt{1-\beta\textbf{E}^2}}
\ee
to obtain
\begin{align}\lb{3.3}
    \frac{\kappa_2^2\textbf{E}^2}{1-\beta\textbf{E}^2}\le\textbf{D}^2\le \frac{\textbf{E}^2}{1-\beta\textbf{E}^2}.
\end{align}
Resolving this inequality, we obtain
\begin{align}\lb{3.4}
    \frac{\textbf{D}^2}{1+\beta\textbf{D}^2}\le\textbf{E}^2\le\frac{\textbf{D}^2}{\kappa_2^2+\beta\textbf{D}^2}\le \frac1\beta.
\end{align}
On the other hand,  the Hamiltonian density \eq{3.5} increases with respect to $\textbf{E}^2$ since it satisfies
\begin{align}
    \frac{{\rm{d}}\mathcal{H}}{{\rm{d}}\textbf{E}^2}=\frac{\kappa_1}{2}+\frac{\kappa_2}{2(1-\beta\textbf{E}^2)^\frac{3}{2}}>0.
\end{align}
Hence, in view of \eq{3.4}, the range of $\mathcal{H}$ is given by
\begin{align}\lb{3.7}
    \frac{\kappa_1\textbf{D}^2}{2(1+\beta\textbf{D}^2)}+\frac{\kappa_2{\bf D}^2}{1+\sqrt{1+\beta{\bf D}^2}}\le\mathcal{H}(\textbf{E}^2)\le\frac{\kappa_1\textbf{D}^2}{2(\kappa_2^2+\beta\textbf{D}^2)}+\frac{{\bf D}^2}{\kappa_2+\sqrt{\kappa_2^2+\beta{\bf D}^2}},
\end{align}
which stays bounded, in particular.

Though the bounds \eq{3.7} are general, it will be of interest to work out a point-charge situation as an illustration. To this end,
Inserting \eq{3.8} into \eq{3.7}, we obtain:
\begin{align}\lb{3.9}
    \frac{\kappa_1q^2}{2\left(\beta q^2+r^4\right)}+\frac{\kappa_2q^2}{r^4+\sqrt{\beta q^2+r^4}\,r^2}\le\mathcal{H}(\textbf{E}^2)\le\frac{\kappa_1q^2}{2(\beta q^2+\kappa_2^2r^4)}+\frac{q^2}{\kappa_2r^4+\sqrt{\beta q^2+\kappa_2^2r^4}\,r^2}.
\end{align}
Thus, using ${\cal H}_1(r)$ and ${\cal H}_2(r)$ to denote the left- and right-hand sides of \eq{3.9},  respectively, we see that the energy $E$ of an electric point charge given by \eq{3.8} lies
in the interval $(E_1,E_2)$ where 
\begin{align}
&\frac{E_{1}}{4\pi}=\int^\infty_0\mathcal{H}_{1}(r)r^2{\rm{d}}r=\int^\infty_0\left(\frac{\kappa_1q^2r^2}{2\left(\beta q^2+r^4\right)}+\frac{\kappa_2q^2}{r^2+\sqrt{\beta q^2+r^4}}\right){\rm{d}}r,\label{3.10}\\
&\frac{E_{2}}{4\pi}=\int^\infty_0\mathcal{H}_{2}(r)r^2{\rm{d}}r=\int^\infty_0\left(\frac{\kappa_1q^2r^2}{2(\beta q^2+\kappa_2^2r^4)}+\frac{q^2}{\kappa_2r^2+\sqrt{\beta q^2+\kappa_2^2r^4}}\right){\rm{d}}r.\label{3.11}
\end{align}
Again, after rescaling, the integrals on the right-hand sides of  \eq{3.10} and \eq{3.11} can be expressed in terms of the quantities
\begin{align}
    \int^\infty_0\frac{x^2{\rm{d}}x}{1+x^4}=\frac{\sqrt{2}\pi}{4},\ \ \int^\infty_0\frac{{\rm{d}}x}{\sqrt{1+x^4}+x^2}=\frac{\pi^{\frac{3}{2}}}{3\Gamma\left(\frac{3}{4}\right)^2},
\end{align}
 such that 
\bea
\frac{E_1}{4\pi}&=&\frac{q^2}a\left(\frac{\sqrt{2}\pi\kappa_1}8+\frac{\kappa_2\pi^{\frac32}}{3\Gamma\left(\frac34\right)^2}\right)\equiv\frac{q^2}a H_1(\kappa_1,\kappa_2),\lb{5.20}\\
\frac{ E_2}{4\pi}&=&\frac{q^2}{a}\left(\frac{\sqrt{2}\pi\kappa_1}{8\kappa_2^{\frac32}}+\frac{\pi^{\frac32}}{3\kappa^{\frac12}_2\Gamma\left(\frac34\right)^2}\right)
\equiv\frac{q^2}a H_2(\kappa_1,\kappa_2).\label{5.21}
\eea
 In view of \eq{3.9}, we see that the normalized energy, $H(\kappa_1,\kappa_2)$,  of the point charge \eq{3.8} enjoys the bounds
\be\lb{5.22}
H_1(\kappa_1,\kappa_2)<H(\kappa_1,\kappa_2)<H_2(\kappa_1,\kappa_2),\quad \kappa_1,\kappa_2>0,
\ee
and the bounds converge to the Born--Infeld result as $\kappa_2\to 1$:
\be
\lim_{\kappa_2\to1} H_1(\kappa_1,\kappa_2)=\lim_{\kappa_2\to1} H_2(\kappa_1,\kappa_2)=H(0,1)=\frac{\pi^{\frac32}}{3\Gamma\left(\frac34\right)^2}.
\ee

\begin{figure}[htbp]
    \centering
    \includegraphics[width=0.5\linewidth]{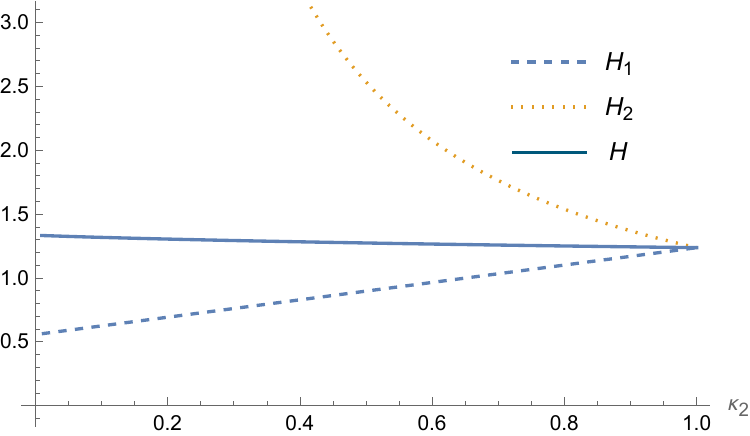}
    \caption{\normalsize{The normalized energy $H(\kappa_1,\kappa_2)$ of an electric point charge in relation to its explicit lower bound $H_1(\kappa_1,\kappa_2)$ and upper bound $H_2(\kappa_1,\kappa_2)$ plotted over the weight parameter $\kappa_2$ showing the convergence of these quantities as $\kappa_2\to1$.}}
    \label{Fig2}
\end{figure}

In Figure \ref{Fig2}, we present a plot of the normalized energy $H$ over the full interval $(0,1]$ of the parameter $\kappa_2$ in comparison with its lower and upper bounds stated in \eq{5.22}. The two bounds converge to the classical Born--Infeld energy as $\kappa_2\to 1$ (hence $\kappa_1\to0$).

\section{Conclusions}
\noindent
In this work, we have formulated an electrodynamic theory 
governed by the action density \eq{model}, where $\cal F$ is either given in \eq{1.1} or more generally given in \eq{4.4}, which is seen as a weighted sum of the Maxwell action density and the Born--Infeld action density
with positive weight parameters $\kappa_1$ and $\kappa_2$ satisfying $\kappa_1+\kappa_2=1$.
 The unique novelty
of this theory is that it accommodates a finite-energy electric point charge as that of the Born--Infeld theory but does not accommodate a finite-energy monopole or a dyon, more
generally, unlike that in the Born--Infeld theory, as a sharp contrast. Specifically, our results may be summarized in the following theorem.

\begin{theorem}
The static equations of motion of the theory always allow a unique solution
describing a finite-energy electric point
charge but never allow a solution giving rise to a finite-energy monopole or dyon. 

Furthermore, these equations can also be solved in the general situation where the electric charge distribution
is arbitrarily given, such that:
(i) The total free electric charge generated from the induced electric field is the same as the total prescribed electric charge associated with the external electric charge density which
determines the electric displacement field.
(ii) The total energy $E$ of an electric point charge $q$ and effective radius $a$ assumes the classical form
\be\lb{6.1}
\frac E{4\pi}=\left(\frac{q^2}a\right)H,
\ee
where $H$ is a constant depending only on the weight parameters, which enables us to estimate the electron radius for arbitrary weight parameters.
(iii) The normalized energy $H$ in \eq{6.1}, hence the total energy of an electric point charge can be calculated exactly in terms of an explicit integral. This energy is a decreasing function of the parameter $\kappa_2$ such that
 a greater Maxwell theory weight renders a higher energy value. Interestingly, when we take the Maxwell theory limit $\kappa_2\to0$ and hence $\kappa_1\to 1$, the energy converges, rather than
diverges, such that the theory produces a finite-energy electric point charge, in particular a finite-energy classical point-charge electron, in the Maxwell theory limit, with the normalized
energy $H$ in \eq{6.1}
determined to be
$\frac43$.
(iv) In the entire parameter range of the theory, the effective radius of electron can also be estimated exactly and explicitly, which is a decreasing function of $\kappa_2$. In
the Maxwell theory limit $\kappa_2\to0$, the ratio of the electron radius against the classical electron radius is determined to be $\frac{20}9$. 
(v) The solutions represent electrostatic solutions to Schwinger's covariant Maxwell equations subject to the Born--Infeld constitutive equations. When the system is that of a
singly-centered electric charge distribution giving rise to radially symmetric electric fields, no magnetic current is present. However, in the situation of a non-singly-centered electric charge
distribution, a magnetic current will be present in order to maintain a non-radially symmetric balance of so-induced electric fields. 

\end{theorem}

In forthcoming works, we will present our results on gravitational consequences of the theory and the results based on some extensions of the theory in the context of generalized
Born--Infeld models.

\medskip

{\bf Data availability statement}: The data that supports the findings of this study are
available within the article.

{\bf Conflict of interest statement}:
The authors declare that they have no known competing financial interests or personal relationships that could have appeared to influence the work reported in this article.

\end{document}